\documentclass[preprint2]{aastex}

\def\url#1{{\ttfamily\def\/{/\discretionary{}{}{}}#1}}

\def\LCDM{{\char'3CDM}}
\def\mathnew{\mathsurround=0pt}
\def\simov#1#2{\lower .5pt\vbox{\baselineskip0pt
    \lineskip-.5pt\ialign{$\mathnew#1\hfil##\hfil$\crcr#2\crcr\sim\crcr}}}  
\def\simgreat{\mathrel{\mathpalette\simov >}}
\def\simless{\mathrel{\mathpalette\simov <}}
\def\'#1{\ifx#1i{\accent"13\i}\else{\accent"13#1}\fi}

\shorttitle{Modeling Dynamical Dark Energy}
\shortauthors{Mainini et al.}
\begin{document}
\title{Modeling Dynamical Dark Energy}

\author{R. Mainini, A.V. Macci\`o \& S.A. Bonometto}
\affil{Physics Department G. Occhialini, Universit\`a degli Studi di
Milano--Bicocca, Piazza della Scienza 3, I20126 Milano (Italy)}
\author{A. Klypin}
\affil{Astronomy Department, New Mexico State University, Box 30001, Department
4500, Las Cruces, NM 88003-0001}

\begin{abstract}

Cosmological models with different types of Dark Energy are becoming
viable alternatives for standard models with the cosmological
constant. Yet, such models are more difficult to analyze and to
simulate.  We present analytical approximations and discuss ways of
making simulations for two families of models, which cover a wide
range of possibilities and include models with both slow and fast
changing  ratio $w=p/\rho$. More specifically, we give
analytical expressions for the evolution of the matter density
parameter $\Omega_m(z)$ and the virial density contrast $\Delta_c$ at
any redshift $z$.  The latter is used to identify halos and to find their
virial masses.  We also provide an approximation for the linear growth
factor of linear fluctuations between redshift $z=40$ and $z=0$.  This is
needed to set the normalization of the spectrum of fluctuations.
Finally, we discuss the expected behavior of the halo mass function
and its time evolution. 
\end{abstract}


\section{Introduction}
Observations of high redshift supernovae \citep{Perlmutter, Riess} 
as well as the
analysis of fluctuations of the cosmic microwave background combined
with data on the large-scale structure of galactic distribution
\citep[e.g.][]{Balbi, Tegmark01, Netterfield, Pogosian03, Spergel03}
 indicate
that there is a significant component of smooth energy with large negative
pressure, characterized by 
a parameter
$w \equiv p/\rho  \simless -0.5$. 
This
component is dubbed dark energy (DE). The nature of DE is open for
debate with candidates ranging from
a
cosmological constant $\Lambda$ to
a slowly evolving
scalar field $\phi$ to even more exotic physics of extra
dimensions \citep[e.g.,][]{DvaliTurner}


One of the most appealing ideas for 
DE
is a self--interacting scalar field, which
evolves with time \citep{RP, wett1}. We call this dynamical Dark
 Energy. The advantage of the dynamical DE models as compared with the
 $\Lambda$CDM models is that DE naturally yield an accelerated
 expansion easing the problem of {\it fine tuning}.  The
 observational signatures of 
dynamical DE should be carefully
 investigated in order to determine which measures can be used to
 discriminate $\Lambda$CDM from dynamical DE and among different
 dynamical DE models.  In this paper we focus on 
the two most popular variants of dynamical DE.
  \citet[][RP hereafter]{RP} studied DE models, which
cause a
 rather slow evolution of $w$.  Models based on simple
 potentials in supergravity (SUGRA) result in 
 faster evolving $w$
 \citep{BraxMartin99, BraxMartin00}.  Together RP and SUGRA potentials
 cover a large spectrum of evolving $w$.  The potentials are written
 as
\begin{eqnarray}
V(\phi) &=& \frac{\Lambda^{4+\alpha}} {\phi^\alpha} ~~~~~~~~\qquad (RP), \\
V(\phi) &=& \frac{\Lambda^{4+\alpha}}{\phi^\alpha} \exp (4\pi G \phi^2)~~~ (SUGRA).
\end{eqnarray}
Here $\Lambda$ is an energy scale, 
currently set in the range $10^2$--$10^{10}\, $GeV, relevant for 
fundamental interaction physics. The potentials depend also on the exponent 
$\alpha$. 
Once the parameters $\Lambda$ and $\alpha$ are assigned, the DE density 
parameter $\Omega_{DE}$ follows. Here, however
we prefer to use $\Lambda$ 
and $\Omega_{DE}$ as independent parameters. 

Dynamical DE has a kinetic and a potential components, reading
$\dot \phi^2/2$ and $V(\phi)$, respectively. 
Those factors
define  the energy density $\rho_{DE}$ and
the pressure $p_{DE}$. In general, the ratio of the pressure and the density
\begin{equation}
   w = {p_{DE} \over \rho_{DE} } = {\dot \phi^2/2  - V(\phi) \over
          \dot \phi^2/2  + V(\phi)}
\end{equation}
changes with time and is typically negative when the potential $V$ is
sufficiently large, as one expects to occur in the recent epoch.  

In order to simplify the situation, the dynamical DE is often replaced
 with models with constant $w \neq -1$. This can be considered as a
 formal generalization of the equation of state of vacuum energy
 density for which $w \equiv -1$. These models result in accelerated
 expansion if $w$ exceeds $\approx -1/3$. 
The main advantage of constant $w$ is to yield models easier to deal with
 than the dynamical DE. 
%
Although
 finding a physical
 justification for models with constant $w \neq -1$ is more difficult than for
 the cosmological constant (see, however, Caldwell 2002),
these models
are still
 useful as toy models, allowing one to inspect the effects of
 an acceleration which is slower than with the vacuum energy.  

In this paper we
show how complications with the dynamical DE can be overcome if one
 uses 
suitable approximations, that we provide.
 Besides of allowing an
 easier treatment of the dynamical DE, 
these expressions
 also allow
 us to compare the dynamical DE with the models with constant $w$.
 One of the results of this comparison is that 
differences between constant--$w$ and dynamical DE are significant,
being comparable with those between $\Lambda$CDM and constant--$w$.

The results given in this paper are based on a modified version of the
CMBFAST code. The modifications include effects due to the change in
the rate of the expansion of the Universe and 
fluctuations of the scalar field. Although these fluctuations
rapidly fade, soon after their enter the horizon, their effect
on cosmic microwave background anisotropies and polarization
is quite significant, while they also cause (smaller) modifications
of the transfer function on large scales.

In addition, we also estimate the growth of linear and non-linear
fluctuations of non--relativistic matter only.  Previously our
algorithms were used by \citet[][MMB03 hereafter]{Mainini03}. 

Making use of these algorithms, in
this paper we 
work out:
(i) Analytical 
approximations
 of the dependence
of the matter density parameter $\Omega_m$ on the redshift $z$; (ii)
Modifications to run $N-$body simulations of clustering of 
dynamical DE models; (iii) Analytical 
approximations
 for the virial density
contrast $\Delta_c$ at any redshift $z$.  Expressions derived from the
linear theory can also be used to compare the observables deduced for
dynamical DE and for constant $w$.  We argue that these
approximations make 
an
analysis of the dynamical DE as simple as for
models with constant $w$.

\section{The virial density contrast}

We start with finding the evolution of the density contrast in the
top-hat 
approximation
 for models with DE.  Considering a spherical fluctuation
greatly simplifies the analytical and numerical treatment of the
non--linear problem. Much work has been done in this line, starting
with Gunn \& Gott (1972), Gott \& Rees (1975) and Peebles (1980), who
studied the spherical collapse in standard CDM (SCDM)
models. Lahav et al. (1991), Eke et al. (1996), Brian \& Norman
(1998) and others generalized the results to the case of the $\Lambda$CDM.
If the initial density contrast of a spherical perturbation is $\Delta_i=
1+\delta_i$ and its initial radius is $R_i$, then the radius of the
perturbation $R=rR_i$ at later times can be found using the equation:
\begin{eqnarray}
{\ddot r \over r} = -H_i^2 \left[ {\Omega_{m,i} \Delta_i \over 2 r^3}
+ \Omega_{r,i} \left(a_i \over a \right)^4 + {(1+3w)\rho_{DE}
\over 2\rho_{cr,i} } \right]
\end{eqnarray}
where all quantities with subscript $i$ refer to the initial time.
In particular $\Omega_{m,i} $ and $\Omega_{r,i} $ are the density
parameters for non- and
 relativistic matter at that time.  After slowing down relative to the
 scale factor $a(t)$, the perturbation eventually stops at moment
 $t_{ta}$, when its radius is $R_{ta}$. The radius $R$ formally goes
 to zero at $\sim 2t_{ta}$ corresponding to redshift $z_{col}$.  The
 value of $z_{col}$ depends on the amplitude of the initial
 fluctuation $\delta_i$. Instead of $\delta_i$ it is however
 convenient to use the amplitude $\delta_c$ as estimated by the linear
 theory at $z_{col}$.  For SCDM the value of this density contrast is
\begin{equation}
    \delta_c^* \simeq 1.68
\end{equation} 
and does not depend on $z_{col}$ (see, e.g., Coles
 \& Lucchin 1995). For other models $\delta_c $ does depend on $z_{col}$ .

In the contraction stages fluctuations heat up and, unless
kinetic energy can be succesfully radiated away, contraction
will stop when virial equilibrium is attained and its size is $R_v$.
Requiring energy conservation and virial equilibrium we obtain
the following algebric cubic equation
\begin{equation}
   x^3 - { 1+y(a_{ta}) \over 2y(a_{col}) } x + {1 \over
            4 y(a_{col})} = 0,
\label{eq:x}
\end{equation}
where $x=R_{v}/R_{ta}$ and
\begin{equation}
   y (a) = {1-\Omega_{m} (a) \over \Delta_i \Omega_m (a) }
       \left( R_{ta} \over R_i \right)^3
       \left( a_{i} \over a \right)^3 ~.
\end{equation}
Note that the actual radius of the final virialized halo is often
larger than $R_v$ 
, owing to deviations from spherical growth in the real world
(Macci\`o, Murante \& Bonometto, 2003).
Still, $R_v$ is
a good starting point for statistical analysis.  Multiplying eq.~(\ref{eq:x}) by
$2y$ and taking then $y=0$ (i.e. $\Omega_m \equiv 1$: SCDM), we
recover that $x=1/2$. In general, the root $x$ lays slightly below
this value.

\begin{figure}[tb!]
\plotone{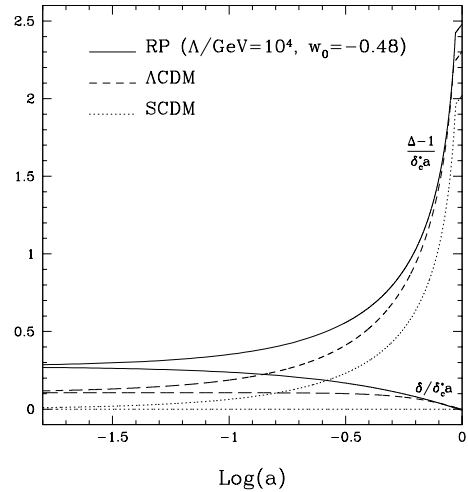}
\caption{\small Normalized linear (bottom curves) and non--linear (top
 curves) amplitude of density fluctuations for SCDM (dotted), \LCDM
 (dashed), and RP (full) models. The amplitude of fluctuation was
 normalized to have collapse if the perturbation at $z_{col}=0$.
 Similar plots can be given for collapse at any other redshift. The
 density contrast $\Delta = \Delta_c/\Omega_m $.  }
\label{fig:delta}
\end{figure}

Figure~\ref{fig:delta} shows the linear and non--linear growth of a
density contrast, for SCDM, $\Lambda$CDM and a RP models normalized
to have $z_{col}=0$.  Similar plots can be made for any redshift of
collapse.  The Figure can be used to find the initial amplitude
$\Delta_i$ at any given redshift $z_i$ and the value of $\delta_c$
for a perturbation collapsing at present.
Using the final value of $\Delta$ we obtain
the virial density contrast:

\begin{equation}
   \Delta_c = \Omega_m \Delta.
\end{equation}

\begin{figure}[tb!]
\plotone{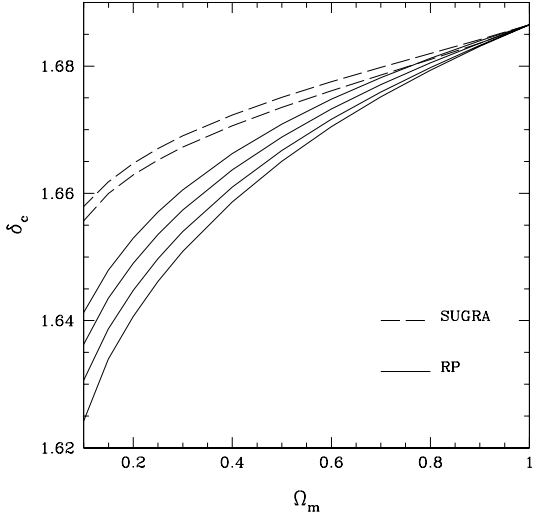}
\caption{\small The dependence of $\delta_c$ on the matter density parameter
$\Omega_m$ at $z=0$  for 4 RP ($\Lambda/$GeV =  $10^2$, $10^4$,
$10^6$ and 10$^8$) and 2 SUGRA models ($\Lambda/$GeV =  $10^2$ and
$10^8$). $\Lambda$ values increase from top to bottom curves.
}
\label{fig:deltac}
\end{figure}

\begin{figure}[tb!]
\plotone{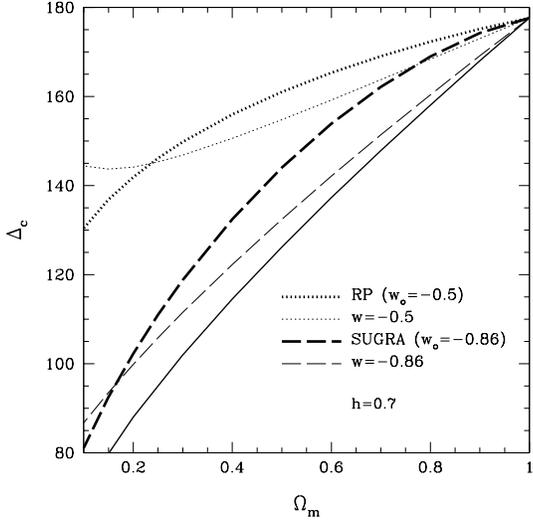}
\caption{\small $\Omega_m$ dependence of $\Delta_c$ for different
cosmologies. RP and SUGRA models, at $z=0$, have a pressure/density
ratio $w_o = w$ of the constant $w$ models shown. The full curve
is for $\Lambda$CDM.
}
\label{fig:Omegam}
\end{figure}

\begin{figure}[tb!]
\plotone{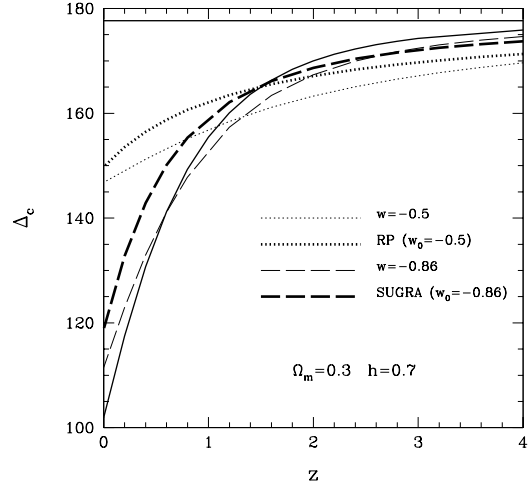}
\caption{\small Redshift dependence of $\Delta_c$ for different
cosmologies. RP and SUGRA models, at $z=0$, have a pressure/density
ratio $w_o = w$ of the constant $w$ models shown. Full curves
are for  DM and $\Lambda$CDM.
}
\label{fig:DeltacZ}
\end{figure}

 In the linear and non--linear cases deviations from the SCDM behavior
often compensate and the final values of $\delta_c$ are just slightly
model dependent (see Figure~\ref{fig:deltac} and MMB03 for more
details).  
The spread among the virial density contrasts $\Delta_c$,
 is  large as indicated by Figure~\ref{fig:Omegam}, which shows $\Delta_c$
as the function of $\Omega_m$ for different models. 
The evolution of $\Delta_c$ with redshift is 
  also very model dependent, as shown by Figure~\ref{fig:DeltacZ}).
  We provide an
approximation,
 which is  valid at any redshift $z$, provided that
we know the matter density parameter $\Omega_m$ at that redshift:
\begin{equation}
   \Delta_c = 178 \, \Omega_m^{\, \mu(\Omega_m,\Lambda)}.
\label{eq:deltac}
\end{equation}
Here $\mu(\Omega_m,\Lambda) = a+b\, \Omega_m^{\, c}$ with $c=1\, (2)$ for 
RP $\, $(SUGRA). Parameters $a$ and $b$ are given by
\begin{equation}
   a=a_1\lambda+a_2, \quad b=b_1\lambda+b_2,
\label{eq:ab}
\end{equation}
where 
\begin{equation}
\lambda = \log(\Lambda/{\rm GeV})
\label{eq:lamlam}
\end{equation}
and the coefficients are given in Table~1.

\begin{table}
\begin{center}                                                             
\caption{Interpolation coefficients for $\Delta_c$} 
 \begin{tabular}{lllll}\hline\hline
 Model  & $a_1$ & $a_2$ &
                $b_1$ &  $b_2$ \\\hline
  RP & $-1.45$ $\times 10^{-2}$ & $0.186$  & $-0.011$   & $0.22$ \\
SUGRA & $-2.25$ $\times 10^{-3}$ & $0.3545$ & $-0.01875$ & $-0.1225$ \\
\hline
\end{tabular}
\end{center}
\label{table:delatc}
\end{table}

Figure~5 shows the dependence on $\lambda$ of the differences
$|\Delta_c^{num}/\Delta_c^{an}-1|$, at $z=0$, for models with $h=0.7$
and different values of $\lambda$, as a function of $\Omega_m$.
(Here $\Delta_c^{num}$ is obtained from the full numerical
treatment, while  $\Delta_c^{an}$ is the expression (\ref{eq:deltac}))
Discrepancies stay below $0.5\, \%$ for any $\Omega_m \simless 0.15$.
However, for large $\lambda$, the approximation is even better:
 $\simless 0.2\, \%$, for any $\Omega_m$.

\begin{figure}[tb!]
\plotone{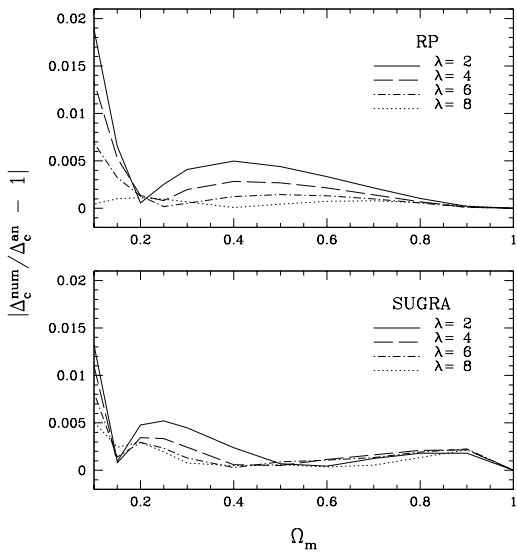}
\caption{\small Fractional discrepancy between numerical and analytical
results on $\Delta_c$.
}
\end{figure}

\section{The mass function and the linear growth factors for dynamical DE}

We use both the Press--Schechter (PS; 1974) and the Sheth--Tormen
(ST; 1999, 2002) approximations for the mass
function of dark matter halos.  The value of $\delta_c$ defines the
bias factor $\nu = \delta_c/\sigma_M$ for the mass $M$. Here
$\sigma_M$ is the rms density fluctuation on this scale. The bias
factor enters then the expression
\begin{equation}
    f(\nu)\nu \, d\log \nu = {M \over \rho_m} n_h(M) M\, d\log M,
\label{eq:massFn}
\end{equation}
with either
\begin{equation}
   f(\nu)\, \nu = \sqrt{2/\pi}\, \,  \nu \exp(-\nu^2/2) ~,
~~~~~~~~~~ {\rm (PS)}   
\label{eq:PS}
\end{equation} or
\begin{eqnarray}
   f(\nu) \, \nu = ~~~~~~~~~~~~~~~~~~~~~~~~~~~~~~~~~~~ & \nonumber \\
 A(1+\nu'\, ^{-2q}) \sqrt{2/\pi}\, \,  
   \nu' \exp(-\nu'\, ^2/2), & {\rm (ST)}~~~
 \label{eq:ST}
\end{eqnarray}
with a small complication in the ST case; here $\nu'=\sqrt{a}\, \nu$
with $a=0.707$, while the constants $q = 0.3$ and $A = 0.3222$.  Using
eq.~(\ref{eq:massFn}) we obtain the differential mass function (MF)
$n_h(M)$ in the PS and ST approximations, once the distribution on
bias is given. Here, as usual, we assume a Gaussian
$f(\nu)$. Eqs.~(\ref{eq:massFn}-\ref{eq:ST}) can then be integrated to
obtain the halo mass function $n_h(>M,z)$ at any redshift $z$.

\begin{figure}[tb!]
\plotone{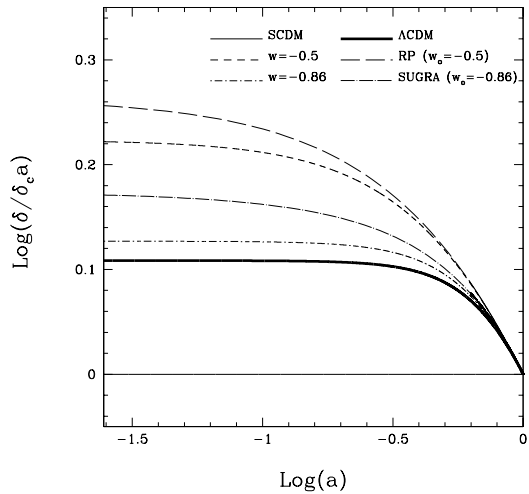}
\caption{\small Linear growth factor for various models. $w_o$ is the
value of $w$ at $z=0$.
}
\label{fig:linear}
\end{figure}

Such computation must use appropriate values for $\delta_c$ and
$\sigma_M$; the latter are computed by integrating the power spectrum
$P(k)$.  Its shape depends on specific choice of dynamical DE as our
modified CMBFAST program shows.  Yet, the dependence is very mild for
wavelength smaller than the galaxy cluster scale. On the contrary, as
can be seen also from Figure~\ref{fig:delta}, the linear growth factor
depends on DE nature in quite a significant way. Figure~\ref{fig:linear} presents the
$z$--dependence of the growth factor for $z$ up to 40 and for a number
of different models.

In particular,  Figure~\ref{fig:linear}  shows that at redshift $z \geq 2$
 the difference between \LCDM ~and a model with 
constant
$w = -0.86$ is equal or even smaller
than the very difference between this constant--$w$ model and the SUGRA model
yielding the same ratio $w$ at $z=0$. However, the latter
difference becomes comparable with the former one already at 
$z \simgreat 0.5$.
A constant--$w$ approximation seems to perform better for RP models,
but this can be mostly ascribed to the fact that the ratio $w$ at $z=0$
is smaller, for these models. Their distance from \LCDM ~is therefore
greater and the difference between them and constant--$w$ models
appears comparatively smaller. However, also in this case,
using constant $w$ instead of RP, at $z \geq 4$ is surely
misleading.

The linear growth factor shown in Figure~\ref{fig:linear} is very important for setting
 initial conditions of $N-$body simulations because linear growth factors are required for
 the normalization of the power spectrum at the initial redshift $z_{\rm in}$ 
of simulations. For these reasons we also give
an analytical approximation which reproduces fairly well the behavior of the
linear growth factors  at $z=40$ for different values of $\Omega_m(z=0)$ and
$\lambda$:
\begin{equation}
   {\delta_c \over \delta(z=40)} = A + B\lambda + C\lambda^2.
\end{equation}
The values of the coefficients $A,B,C$ are presented in Table~2 
for RP and SUGRA models respectively.
At $z=40$, the discrepancies between $\Omega_m$ and unity already
range around 2--3$\, \%$. If a simulation must be started at larger $z$,
extrapolating the linear growth factor by assuming that $\delta \propto a$
at $z>40$, implies an error smaller than such percentage. This can be still
improved by assuming that $\delta \propto a^{\Omega_m^q}$,
with $q \simeq 0.4$. The dependence of $q$ on the model and on
the energy scale $\lambda$ fixes the second decimal of $q$ and allows
a precision better than 0.01$\, \%$, which is out of the scopes of
this analysis.
%

\begin{table}
\begin{center}  
\caption{Coefficients for the linear growth factor}
 \begin{tabular}{lllll}\hline\hline
 Parameter &  $\Omega_m=0.2$ & $\Omega_m=0.3$ &
                        $\Omega_m=0.4$   \\ \hline
SUGRA \\
 $A$    &   25.6 & 28.5 & 30.7 \\
 $B$    &  $-0.237$ & $-0.26$ & $-0.274$ \\
 $C$    &   0 & 0 & 0 \\
\\
RP \\
 $A$    &   21.3 & 25.1 & 28.2 \\
 $B$    &  $-0.755$ & $-0.783$ & $-0.698$ \\
 $C$    &   $-0.0125$ & $-0.0155$ & $-0.0155$ \\
\hline
\end{tabular}
\end{center}
\label{table:lgf}
\end{table}

Mass functions
$n(>M,z)$, obtained according to eqs.~(\ref{eq:massFn}-\ref{eq:ST}), will be compared 
with simulations in the accompanied paper (Klypin, Macci\`o, Mainini
\& Bonometto 2003).
Similar mass functions, obtained from the PS expressions, were
used in MMB03 to estimate expected observable
differences between models with different dynamical DE.

\section{Evolution of the matter density parameter}

In RP and SUGRA models, at variance from models with $w = {\rm
const}$, no analytical expression of $\Omega_m(a)$ is readily
available.  An accurate approximate expression of $\Omega_m$, for
various redshifts and for different models is useful for various
purposes.  In particular, it can be used, in conjunction with
eqs.~(\ref{eq:deltac}-\ref{eq:ab}), to find the value of
$\Delta_c$ at $z \neq 0$.

We found the following fitting formula:
\begin{equation}
   \Omega_m(a) = 1-(1-\Omega_{m,0})/(1+z)^{\alpha(z,\lambda)},
\label{eq:OmegA}
\end{equation}
where $\Omega_{m,0}$ is the matter density parameter at $z=0$, while
$ \alpha(z,\lambda) = a + bz^c + d/(1+z) $ with $d=0$ for RP models.
Parameters $a$, $b$, $c$ and $d$ have the same structure as eq.~\ref{eq:ab}.
The coefficients are given in Tables~3.

\begin{deluxetable}{lllllll}
\tablecolumns{8}
\tablewidth{0pc} 
\tablecaption {Coefficients for $\Omega_m(z)$}
\tablehead{ 
\colhead{Parameter} & \colhead{$\Omega_m=0.2$} &  \colhead{$\Omega_m=0.3$}
&\colhead{$\Omega_m=0.4$}\\}
\startdata
RP \\
 $a_1$    &  $-5.638\times10^{-3}$ & $-2.119\times10^{-2}$ 
          & $-3.365\times10^{-2}$ \\
 $a_2$    &  $-0.813$ & $-0.259$  &  $0.207$   \\
 $b_1$    &  $-2.460\times10^{-2}$ & $-1.833\times10^{-2}$ 
	  & $-1.384\times10^{-2}$ \\
 $b_2$    &   $1.382$	& $0.975$	   &  $0.628$ \\
 $c_1$    &  $-5.960\times10^{-3}$ & $-6.975\times10^{-3}$ 
	  & $-8.394\times10^{-3}$ \\
 $c_2$    &   $8.460\times10^{-2}$  &  $9.771\times10^{-2}$ &  $0.119$ \\
\\
SUGRA \\
 $a_1$    & $-8.466\times10^{-3}$ & $-9.161\times10^{-3}$ & $-2.035\times10^{-2}$ \\
 $a_2$    &  $1.383$	   &  $1.415$     &  $1.427$  \\
 $b_1$    & $-1.386\times10^{-2}$ & $-1.753\times10^{-2}$ & $-1.336\times10^{-2}$ \\
 $b_2$    & $-8.521\times10^{-3}$ & $-6.890\times10^{-3}$ & $-1.289\times10^{-2}$ \\
 $c_1$   & $-3.935\times10^{-2}$ & $-4.421\times10^{-2}$ & $-4.203\times10^{-2}$  \\
 $c_2$     &  $0.710$     &  $0.688$     &  $0.682$	\\
 $d_1$  & $2.088\times10^{-2}$ &  $1.875\times10^{-2}$ & $2.212\times10^{-2}$ \\
 $d_2$   & $-0.883$     & $-0.621$     & $-0.416$	\\
\enddata
\label{tab4}
\end{deluxetable}


Figure~7 shows the errors of approximation
$|\Omega_m^{num}/\Omega_m^{an}-1|$ as a function of the redshift $z$
for two RP and for two SUGRA models with $\Omega_m = 0.3$ and $h=0.7$.

\begin{figure}[tb!]
\plotone{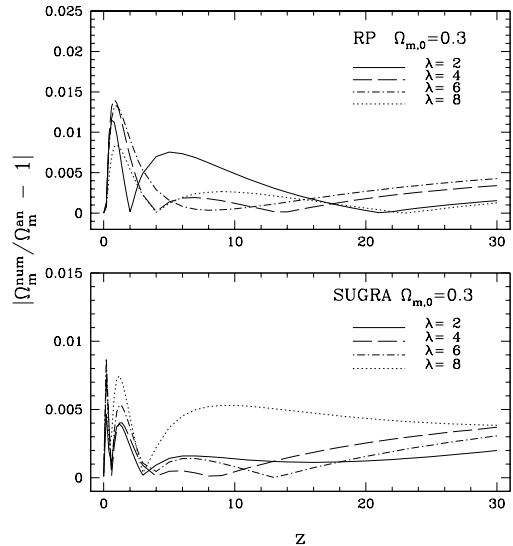}
\caption{\small Fractional discrepancies between the approximated
expression (\ref{eq:OmegA}) and numerical data. 
}
\end{figure}

All that is needed to find the relation between the scale factor $a$ and
time in any flat dynamical DE model, is such expression. In fact, let us remind that
\begin{equation}
    \dot a/ a  = H_o \sqrt{\rho(a)/ \rho_{cr,0}}
\label{eq:Hubble}
\end{equation}
with
\begin{equation}
   \rho(a) = \rho_{m,0}/a^3 + \rho_r/a^4 + \dot \phi^2/2 + V(\phi)~.
\end{equation}
At low $z$, we can omit the contribution of the radiation
density. Therefore, at any time, 
\begin{eqnarray}
   \rho_\phi &=& {\dot \phi^2 \over 2} + V(\phi) = \rho_{cr}(a) [1-\Omega_m(a)]
     \nonumber \\
    &=& \rho_m(a) {1-\Omega_m(a) \over \Omega_m(a)}~,
\end{eqnarray}
provided that we are dealing with a  model, such that the
total density is equal to the critical density $\rho_{cr}(a)$. Then, the Friedmann
equation reads
\begin{eqnarray}
   \big(\, {\dot a \over a}\, \big)^{\, 2} &=& {8 \pi \over 3}G \rho_m(a) \, 
   \big[\,  1 + {1-\Omega_m(a) \over \Omega_m(a)}\,  \big] \nonumber \\
&=&     {8 \pi \over 3}G {\rho_{m,0} \over a^3 \Omega_m(a)} ~,
\end{eqnarray}
so that
\begin{equation}
   {\dot a \over a} = H_o \sqrt{\Omega_{m,0} \over a^3 \Omega_{m} (a)}~.
\label{eq:crd}
\end{equation}
This formula is valid regardless of the equation of state of DE.  In
models with constant $w$, the density $\rho_{DE} \propto a^{-3(1+w)}$
and, therefore, owing to eq.~(21),
\begin{equation}
   \Omega_m(a) = [1+a^{-3w}(\Omega_{m,0}^{-1} - 1)]^{-1} ~.
\label{eq:OmegW}
\end{equation}

The expressions (\ref{eq:OmegA}--\ref{eq:OmegW}), yielding $\Omega_m(a)$,
as well as eq.~(\ref{eq:crd}), yielding $\rho_{cr}$, can be used in 
$N-$body programs, to determine the trajectories of particles in an 
expanding universe. In fact, once we know $\Omega_m$
and $\rho_{cr}$, we can integrate the Poisson equation $\nabla^2 \Phi =
- 4\pi G a^2 \rho_{cr} \Omega_m \delta_m$, yielding the peculiar
potential $\Phi$ due to the density fluctuations $\delta_m$, obtained 
from the particle distribution. Then, the equations of motion of each
particle 
\begin{equation}
   {d \vec p \over da} = - \dot a 
\nabla_x \Phi,  
~~~~~~~~~{d \vec x \over da} = 
   {\vec p  \over a^2 \dot a}
\end{equation}
(see, e.g., Peebles 1980; here  $\vec p \equiv a\vec v$) can be 
integrated, using $\dot a$ given by eq.~(\ref{eq:crd}), and we obtain the 
evolution of particle positions, as a function of the scale factor $a$.
The $N-$body code ART (Kravtsov et al. 1997), used
in the accompanying paper (Klypin, Macc\`o, Mainini \& Bonometto 2003)
to discuss the evolution of models 
with the dynamical DE and DE with constant $w$, has been modified
on these bases.

Figure~8 compares the expansion law $a_{\rm apx}(t)$, obtained using
eqs.~(\ref{eq:OmegA}) and (\ref{eq:Hubble}), with the numerical
behavior $a_{\rm num}(t)$.  Discrepancies seldom exceed $0.4\%$ and
mostly are well below $0.1\, \%$. For any practical reasons the errors
are negligible.


\begin{figure}[tb!]
\plotone{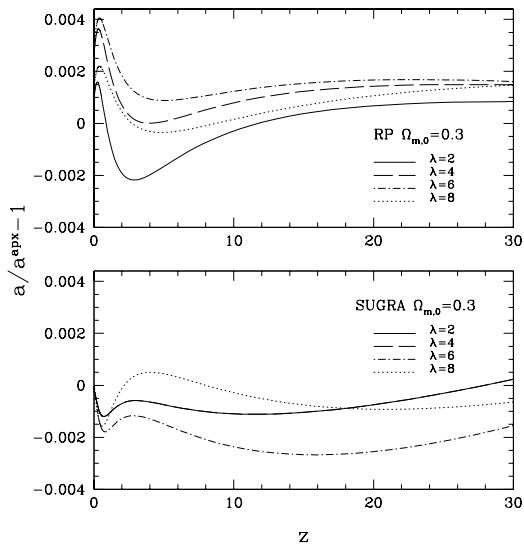}
\caption{\small Fractional discrepancies between the analytical 
and numerical integration of eq.~(\ref{eq:Hubble}) to obtain $a(t)$
}
\end{figure}


\section{Discussion}

Observational effects of DE have been considered by various authors,
but often models with a constant $w$ are used. Besides of being
simpler, constant $w$ models give a feeling that results are generic in
a sense that they do not depend on the nature of underlying dark
energy.  For instance, Wang \& Steinhardt (1998), Steinhardt, Zlatev
\& Wang (1999), Zlatev, Wang \& Steinhardt (1999) and Lokas (2002)
derived the $\Delta_c$ dependence on $\Omega_m$ and $w$, in the
constant $w$ approximation. Schuecker et al (2003) extended the
results to large negative $w$ values to include the case of {\it
phantom energy} (Caldwell 2002, Schulz \& White 2001) .

\begin{figure}[tb!]
\plotone{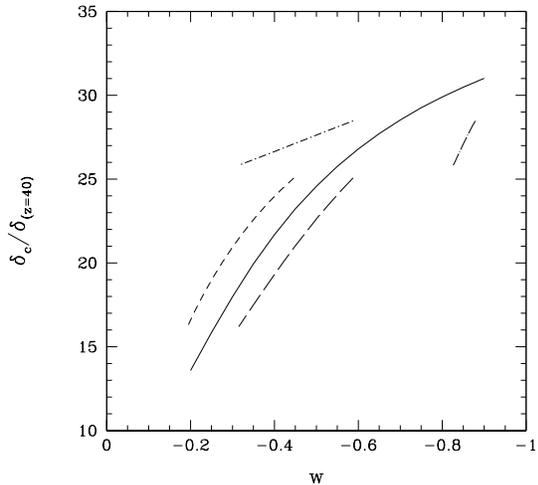}
\caption{\small The linear growth factor for models with
$w= {\rm const}$ (dotted curve) is compared with linear wrowth factor 
for RP and SUGRA models (dashed and dot--dashed curves, respectively). 
RP and SUGRA results are plotted as a function of the value of $w$
they have either at $z=0$ or at $z=40$ (long and short dashes,
respectively). The logarithmic energy scale $\lambda$, for both models,
ranges here from 2 to 10. The plot illustrates that the growth factor for
dynamical DE cannot be easily approximated by any model with 
constant $w$.  }
\label{fig:growth}
\end{figure}

Unfortunately, results depend on what is assumed for the DE.  Figure~3
shows the dependence of $\Delta_c$ on $w$ for models with $\Omega_m =
0.3$ and $h=0.7$ for three cases: DE is cosmological constant,
constant $w \neq -1$, and for dynamical DE with RP or SUGRA
potentials. The difference between constant $w$ and dynamical DE is as
large as the difference between \LCDM ~and a constant $w$. In other
words, if we need to consider models more sophisticated than \LCDM ,
it is not enough to discuss only constant $w \neq
-1$. Figure~\ref{fig:growth} illustrates that the growth factor for
dynamical DE cannot be approximated by a model with constant $w$.  It
seems clear that the Universe ``knows'' the underlying physics, and
predictions depend on the shape of the potential of the scalar field
responsible for the DE.

In principle, finding astrophysical quantities which depend on microphysics
is far from being unwelcome. Accordingly, the detailed dependence
of astrophysical observables on microphysical parameters deserves to be
inspected. Let us outline, in particular, that Figure~3 applies to 
observations at $z=0$, while effects of dynamical DE  are also expected
at higher $z$. In fact, in Figure~4 we show the $z$ dependence 
of $\Delta_c$, for 3 sets of models characterized by the same values 
of $w$ at $z=0$. The figure shows that the differences between $\Delta_c$'s
increase from $z=0$ toward intermediate redshifts, to go back
to SCDM values at high redshifts, when ordinary matter
gradually approaches critical density. However, at intermediate redshifts,
which are most relevant for present and future observations
and for the actual dependence on the nature of DE, at these $z$, arises from
actual changes of $w$, as shown in Fig.~9. 
Apart of any consideration concerning fundamental physics,
therefore, high--$z$ $\Delta_c$ values obtained within constant $w$ approximation risk to create a bias.

\begin{figure}[tb!]
\plotone{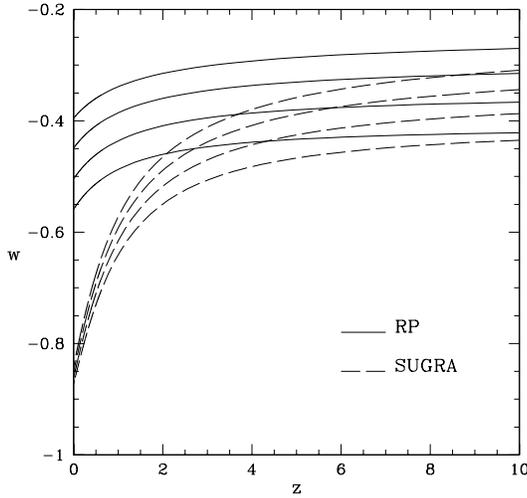}
\caption{\small Redshift dependence of $w$ for 4 RP and 4 SUGRA models 
($\lambda$ = 1, 3, 5 and 7); $\lambda$ decreases 
from top to bottom curves.
}
\end{figure}

Our aim is to facilitate the usage of dynamical DE. 
We provide the following tools: (i) An approximation 
 for $\Omega_m(a)$; (ii) An interpolating 
expression for $\Delta_c$, valid
at any redshift for given $\Omega_m(a)$; (iii) An analytical 
expression for the rate of change of the expansion parameter  
needed for running $N-$body and hydro- simulations; 
(iv) A plot of the linear growth factor, for a number of dynamical DE models, 
and an analytical approximation for it,
to be used to set the initial conditions of $N-$body simulations.

Using these formulae, we modified the ART code, that will be used 
in the accompanying paper (Klypin, Macci\`o, Mainini
\& Bonometto 2003). In a similar way, other programs dealing
with $N-$body interactions or hydrodynamics
can be appropriately modified.

\end{document}